# Anticipated Performance of the
# Square Kilometre Array - Phase 1 (SKA1)

Version 1.0
Robert Braun, Anna Bonaldi, Tyler Bourke, Evan Keane, Jeff Wagg
*SKA Organisation, Jodrell Bank, Lower Withington, Macclesfield, Cheshire SK11 9FT, UK*

**Summary:**
The Square Kilometre Array (SKA), currently under design, will be a transformational facility for studying the Universe at centimetre and metre wavelengths in the next decade and beyond. This paper provides the current best estimate of the anticipated performance of SKA Phase 1 (SKA1), using detailed design work, before actual on-sky measurements have been made. It will be updated as new information becomes available. The information contained in this paper takes precedent over any previous documents.

## 1   Introduction

The SKA Observatory will construct and operate transformational radio telescope arrays with the primary goal of providing major breakthroughs to questions regarding the history of the Universe, primarily through observations of its most abundant constituent, hydrogen. In addition, the SKA will play a major role in addressing key questions in modern astrophysics and cosmology; these are laid out in detail in the most recent SKA science books, "Advancing Astrophysics with the SKA" (AASKA2015; see Braun et al. 2015 for a summary), and are not repeated here.

Due to the challenges in designing and building such large telescopes, SKA is being built in two phases, with Phase 1 consisting of two telescopes, SKA1-Mid and SKA-Low, covering 50 MHz to at least 15 GHz. This paper provides the current best estimates of the anticipated science performance of SKA1, before any actual on-sky measurements are available. SKA1 is currently going through its design phase, with construction to begin in 2021, and so the estimates in this paper are based on the information contained in the design documents and limited prototyping presently available. As this information is updated, and as on-sky measurements become available, this document will be updated to provide the most current performance numbers. The information in this document supersedes previous performance estimates for SKA1, in particular those presented in Dewdney et al. (2013) and summarised in Huynh & Lazio (2013),

The SKA telescopes will be located on two radio quiet sites, in South Africa, home to the MeerKAT and HERA (Hydrogen Epoch of Reionization Array) telescopes, at an altitude of ~1000 m) and Australia, home to the ASKAP (Australia SKA Pathfinder) telescope and the Murchison Widefield Array (MWA); the SKA headquarters are located at the Jodrell Bank Observatory in the UK.

The Design Baseline (not to be confused with an interferometer baseline) identifies the design choices for SKA1 and has been described in previous documents (Dewdney et al. 2013, 2016, 2019) and has evolved during the current design phase. SKA1-Mid will consist of 133 15-m SKA1 offset-Gregorian dishes (Figure 1) combined with the existing 64 13.5-m

MeerKAT dishes, covering 350 MHz to at least 15 GHz, with baselines up to 150 km. SKA1-Low will consist of 131,072 log-period dipole antennas, grouped into 512 stations of size 35-40 m each containing 256 antennas, covering 50 to 350 MHz with baselines up to 65 km. The antennas in an SKA1-Low station work together to act in an analogous manner to a dish (Figure 2).

The frequency coverage and bandwidth of the telescopes is listed in Table 1. While the frequency coverage of the Design Baseline for SKA1-Mid stops at 15.3 GHz, the dishes have an aperture efficiency specification for 20 GHz and as will be evident later in this document there is the expectation that good performance will be possible up to at least ~25 GHz for possible future expansion in frequency space.

**Table 1.** Frequency coverage of SKA1 in the Design Baseline. Bands listed in bold will be deployed as part of the funded Design Baselines. While Bands 3 and 4 are part of the Design Baseline they are not funded at present.

| SKA1 Band | Frequency Range | Available Bandwidth |
|---|---|---|
| **Low** | 50 – 350 MHz | 300 MHz |
| **Mid Band 1** | 0.35 – 1.05 GHz | 700 MHz |
| **Mid Band 2** | 0.95 – 1.76 GHz | 810 MHz |
| Mid Band 3 | 1.65 – 3.05 GHz | 1.4 GHz |
| Mid Band 4 | 2.80 – 5.18 GHz | 2.38 GHz |
| **Mid Band 5a** | 4.6 – 8.5 GHz | 3.9 GHz |
| **Mid Band 5b** | 8.3 – 15.3 GHz | 2 x 2.5 GHz |

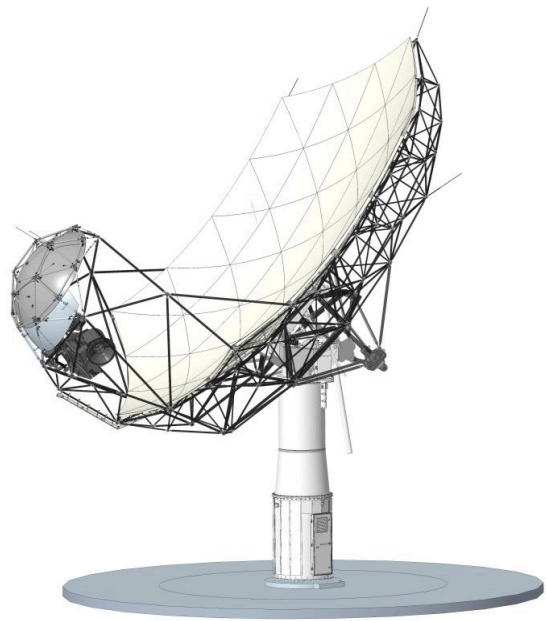

**Figure 1** SKA1-Mid Dish Concept. The proposed dish for SKA is a 15.0-meter multi-panel dual shaped offset Gregorian antenna. The antenna reflector is configured with the feed arm, feed indexer and passive sub reflector

below the main reflector (feed down concept). Several feeds will be integrated into the feed indexer, which allows precision positioning of a single feed at the secondary focus of the ellipsoidal subreflector.

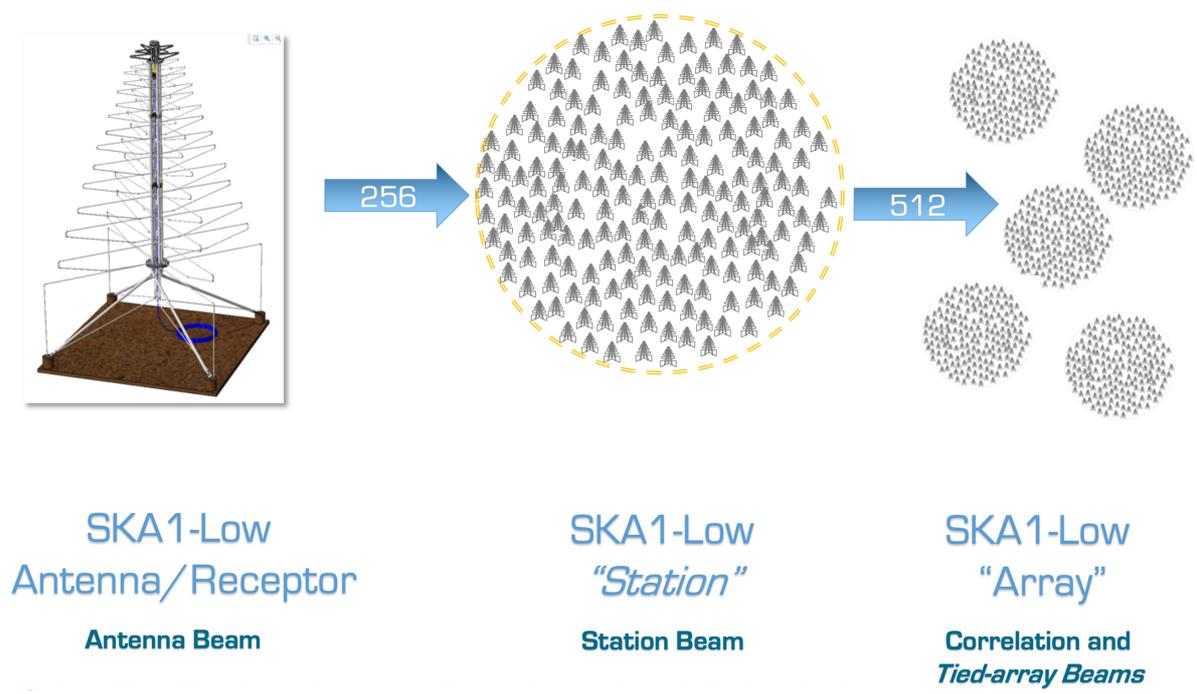

**Figure 2** Illustration of SKA1-Low Antenna, Station, and Array concepts.

Note that Table 1 lists the frequency coverage of the feeds on the SKA1 dishes; the frequency coverage of the MeerKAT dishes is not the same as they have a different feed system. Although the MeerKAT receivers will be retro-fitted to be the same as SKA1, the feeds will not, at least initially. The MeerKAT frequency coverage is given in Table 2. The MeerKAT dishes are also expected to have good performance at higher frequencies, as it was planned to deploy X-band (8-12 GHz) feed/receiver systems on them.

**Table 2.** Frequency coverage of the deployed feed/receiver systems on MeerKAT.

| MeerKAT Band | Frequency Range (GHz) |
|---|---|
| UHF Band | 0.58 – 1.02 |
| L Band | 0.90 – 1.67 |
| S Band | 1.65 – 3.05 |

A summary of the anticipated imaging performance of SKA1 is given in Table 3 with more detail given in the following sections.

**Table 3** Summary of anticipated imaging performance of SKA1

| Nominal Frequency | 110 MHz | 300 MHz | 770 MHz | 1.4 GHz | 6.7 GHz | 12.5 GHz |
|---|---|---|---|---|---|---|
| **Range [GHz]** | 0.05-0.35 | 0.05-0.35 | 0.35-1.05 | 0.95-1.76 | 4.6-8.5 | 8.3-15.3 |
| Telescope | Low | Low | Mid | Mid | Mid | Mid |
| FoV [arcmin] | 327 | 120 | 109 | 60 | 12.5 | 6.7 |
| Max. Resolution [arcsec] | 11 | 4 | 0.7 | 0.4 | 0.08 | 0.04 |
| Max. Bandwdith [MHz] | 300 | 300 | 700 | 810 | 3900 | 2 x 2500 |
| Cont. rms, 1 hr [µJy/beam][a] | 26 | 14 | 4.4 | 2 | 1.3 | 1.2 |
| Line rms, 1 hr [µJy/beam][b] | 1850 | 800 | 300 | 140 | 90 | 85 |
| Resolution Range for Cont. and Line rms [arcsec][c] | 12–600 | 6–300 | 1–145 | 0.6–78 | 0.13–17 | 0.07–9 |
| Channel width (uniform resolution across max. bandwidth) [kHz] | 5.4 | 5.4 | 13.4 | 13.4 | 80.6 | 80.6 |
| Spectral zoom windows x narrowest bandwidth [MHz] | 4 x 3.9 | 4 x 3.9 | 4 x 3.1 | 4 x 3.1 | 4 x 3.1 | 4 x 3.1 |
| Finest zoom channel width [Hz] | 226 | 226 | 210 | 210 | 210 | 210 |

a. Continuum sensitivity at Nominal Frequency, assuming fractional bandwidth of $\Delta\nu/\nu = 0.3$
b. Line sensitivity at Nominal Frequency, assuming fractional bandwidth per channel of $\Delta\nu/\nu = 10^{-4}$ ($>10^{-6}$ will be possible)
c. The sensitivity numbers apply to the range of beam sizes listed
For more details refer to the sections below and Braun et al. 2017, "Anticipated SKA1 Science Performance" (SKA-TEL-SKO-0000818)

# 2 Sensitivity Calculations

## 2.1 SKA1-Mid

### 2.1.1 Sensitivity of an SKA1 dish

The sensitivity of a dish can be expressed as the effective collecting area, $A_{eff}$, divided by the total system temperature, $T_{sys}$. We refer to this as the Sensitivity Metric, $S_M$.

$$S_M = A_{eff}/T_{sys} \ [m^2/K]$$

The effective collecting area can be expressed as,

$$A_{eff} = A_{phys} \ \eta_A,$$

the product of the physical antenna aperture $A_{phys}$ with an aperture efficiency, $\eta_A$. The aperture efficiency can in turn be expressed as the product of several contributing factors, most notably the feed illumination efficiency, $\eta_F$, the phase efficiency, $\eta_{ph}$, of the reflector surface, and a term to capture the large diffractive losses at the lowest frequencies where the reflectors might only be a small number of wavelengths in diameter, $\eta_D$, giving

$$\eta_A = \eta_F \ \eta_{ph} \ \eta_D.$$

In general, all of these factors will vary with frequency.

Extensive simulations and a smaller number of prototype measurements are presented in the design documentation for the dishes (Billade & Dehlgren 2018; Lehmensiek 2018; Leech et al. 2018; Peens-Hough 2018).

A simple empirical model for $\eta_F$ and $\eta_D$ that captures the dependencies and provides numerical agreement over the frequency ranges of relevance is

$$\eta_F = 0.92 - 0.04 \ | \log_{10}(\nu_{GHz}) |$$
$$\text{and}$$
$$\eta_D = 1 - 20 \ (\lambda/D)^{3/2},$$

in terms of the observing frequency in GHz, $\nu_{GHz}$, the wavelength, $\lambda$, and the dish diameter, D.

The phase efficiency can be expressed as

$$\eta_{ph} = \exp(-\Delta_{ph}^2),$$

where

$$\Delta_{ph} = 2\pi\delta/\lambda,$$
$$\text{and}$$
$$\delta = 2(A_p \ \varepsilon_p^2 + A_s \ \varepsilon_s^2)^{1/2}.$$

The anticipated RMS surface errors of the primary and secondary reflector surfaces, $\varepsilon_p = 280$ μm and $\varepsilon_s = 154$ μm, as well as the constants appropriate for the design optics, $A_p = 0.89$ and $A_s = 0.98$, are provided in Lehmensiek (2015).

The system temperature can be expressed approximately as,

$$T_{sys} = (T_{rcv} + T_{spl} + T_{sky})_x / \exp[-\tau_0 \sec(z)],$$

in terms of the receiver temperature, $T_{rcv}$, the spill-over temperature, $T_{spl}$, the sky temperature, $T_{sky}$, the atmospheric zenith opacity, $\tau_0$, and the zenith angle, z. The subscript "x" to the temperature term is used to indicate that a correction of the form $T_x = T \{(h\nu/kT)/[\exp(h\nu/kT) - 1]\}$ is applied.

The sky temperature can be expressed as

$$T_{sky} = T'_{cmb} + T'_{gal} + T_{atm},$$

the sum of CMB, Galactic and atmospheric contributions. $T_{cmb}$ is simply 2.73 K.

$T_{gal}$ depends on the pointing direction of the telescope and can be approximated, away from the Galactic plane, where thermal free-free emission is minimal, with

$$T_{gal} = T_{408} (0.408/\nu_{GHz})^{2.75} \text{ K}.$$

The leading constant has values of $T_{408} = 17.1$, 25.2 and 54.8 K for the 10th, 50th and 90th percentile of the all-sky distribution, when gridded in an equal area projection. The primed quantities, $T'_{cmb}$ and $T'_{gal}$, are attenuated by the atmospheric absorption, $\exp[-\tau_0 \sec(z)]$. The atmospheric contribution depends on the site and weather attributes, specifically the site elevation, atmospheric pressure and zenith water vapour content, together with the zenith angle of an observation. We use the ATM code (Pardo et al., 2001) as implemented within CASA with its default atmospheric profile parameters, together with a representative elevation (near the core of SKA1-Mid) of 1100 m, ambient temperature of 290 K and atmospheric pressure of 895 mbar to calculate the zenith atmospheric emission and opacity at the SKA1-Mid site at three indicative values of the precipitable water vapour (PWV) of 5, 10 and 20 mm. These are shown in Figure 3 and Figure 4.

Weather statistics from the European Centre for Medium-Range Weather Forecasts (ECMWF) for the SKA1-Mid site over the interval 2010 – 2015 have been used to estimate the PWV as function of month of the year by Peter Forkman and John Conway using the methodology described in Ashawaf et al. (2017) and the 10th, 50th and 90th percentiles of those distributions are plotted in Figure 4 together with the same curves for the VLA site. The yearly average values of P10, P50 and P90 for the SKA1-Mid are 5.8, 10.7 and 19.2 mm. For comparison, we note that for the VLA site these are 5.7, 9.6 and 14.5 mm.

**Table 4.** Comparison of average PWV for SKA1-Mid and VLA sites.

| | Yearly Average PWV (mm) | | |
|---|---|---|---|
| **Percentile** | **10th** | **50th** | **90th** |
| **SKA1-Mid** | 5.8 | 10.7 | 19.2 |
| **VLA** | 5.7 | 9.6 | 14.5 |

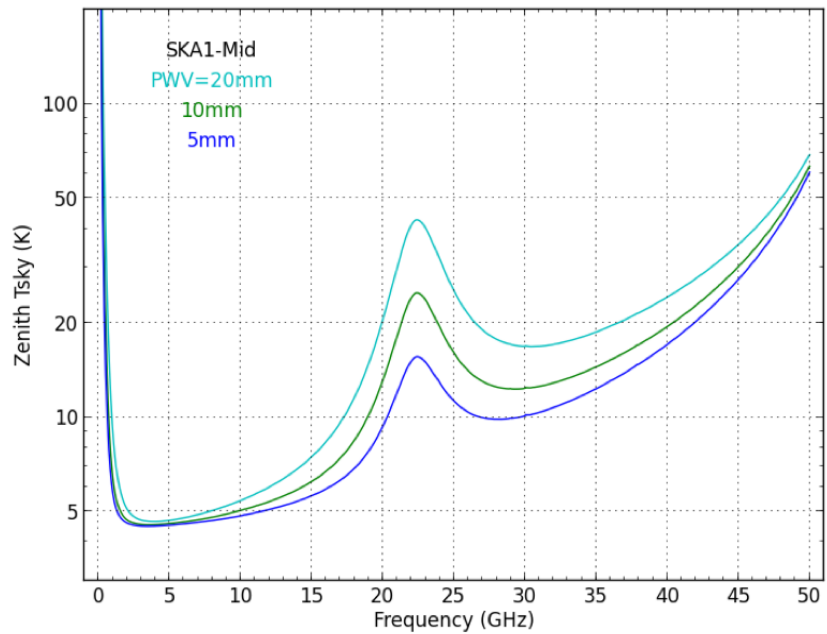

**Figure 3**. Zenith $T_{sky}$ at the SKA1-Mid site for PWV = 5, 10 and 20 mm contributions to the atmospheric emission and the 10th, 50th and 90th percentiles of the Galactic foreground brightness. Atmospheric emission is based on the ATM code (Prado et al. 2001) as embedded in CASA.

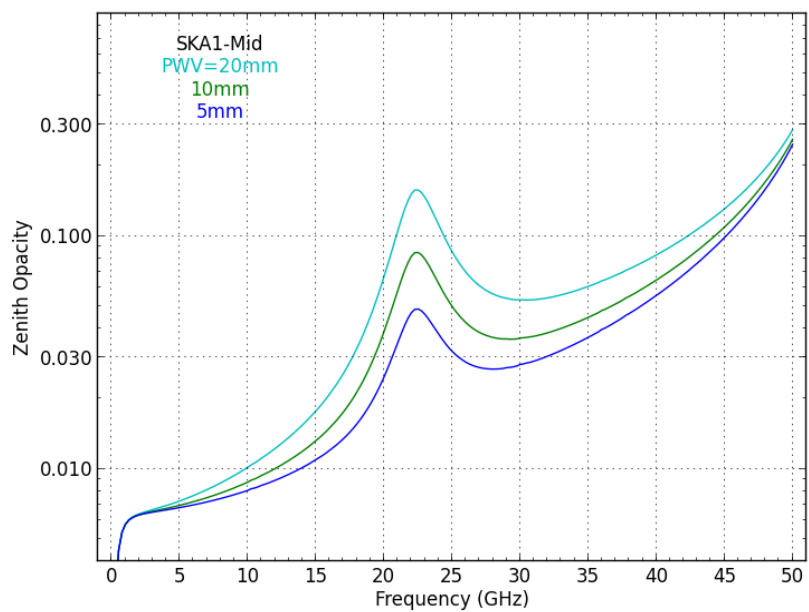

**Figure 4.** Zenith opacity at the SKA1-Mid site for PWV = 5, 10 and 20 mm contributions to the atmospheric emission. Atmospheric absorption is based on the ATM code (Prado et al. 2001) as embedded in CASA.

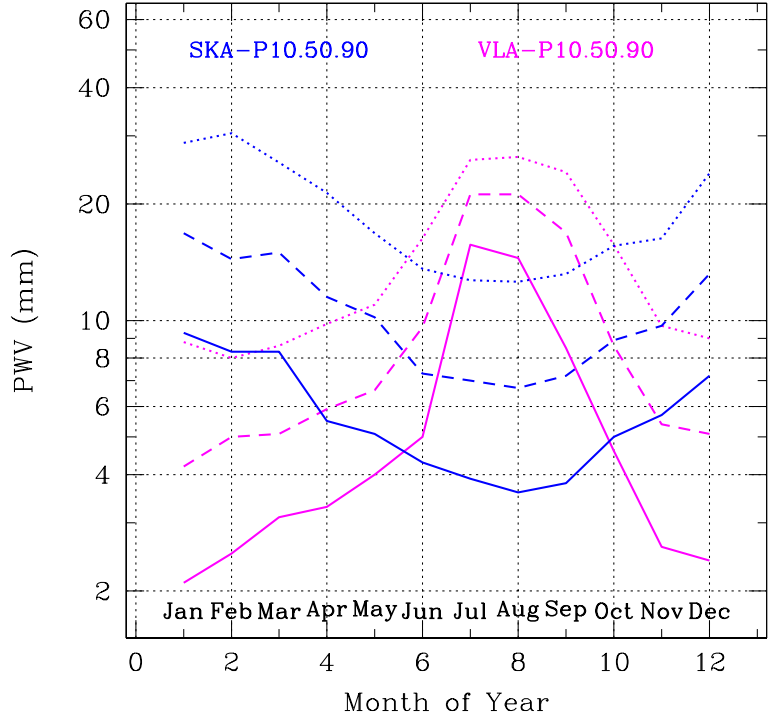

**Figure 5.** Average PWV over the period 2010 – 2015, as derived by Forkman and Conway after Alshawaf et al. (2017), for the SKA1-Mid and VLA sites. The 10[th], 50[th] and 90[th] percentiles of the PWV distribution are plotted against month of the year.

The SKA offset Gregorian shaped dish design, with its "feed down" geometry and extended secondary reflector provides very low spill-over temperatures for zenith angles within about 45 degrees of zenith. Current indications are that the optimised "octave band" feeds being deployed at 950 MHz and above should provide $T_{spl} \approx 3$ K. The 3:1 bandwidth Quad-Ridge Feed Horn design being used for Band 1 (350 – 1050 MHz) is likely to have a higher $T_{spl}$, but for simplicity we will assume a constant value of $T_{spl} = 3$K at all frequencies and capture the documented variation of sensitivity performance with frequency within Band 1 in our model of $T_{rcv}$.

We model $T_{rcv}$ for each of the SKA1-Mid bands guided both by the published $T_{rcv}$ models within the Dish Preliminary Design Review (PDR) documentation, as well as the simulations that yield the final values of $A_{eff}/T_{sys}$ as documented in Billade & Dehlgren (2018), Lehmensiek (2015, 2018), Leech et al. (2018), & Peens-Hough (2018). For the various bands, this yields:

Band 1 (0.35 – 1.05 GHz): $T_{rcv} = 15 + 30 (\nu_{GHz} – 0.75)^2$ K
Band 2, 3, 4 (0.95 – 4.6 GHz): $T_{rcv} = 7.5$ K
Band 5+ (4.6 – 50 GHz): $T_{rcv} = 4.4 + 0.69 \nu_{GHz}$ K

As already noted, the $T_{rcv}$ model, particularly for Band 1, is used to capture several effects in order to reproduce the appropriate $A_{eff}/T_{sys}$. The model at higher frequencies (> 16 GHz) is deliberately conservative, as it represents an extrapolation beyond what has received study. For comparison, Bob Hayward of NRAO has shown that the model $T_{rcv} = 2 + 0.5\nu_{GHz}$ K represents a plausible target for $\nu_{GHz} = 4 – 50$ GHz (Hayward 2012).

The combined model for the sensitivity of a single SKA1 dish is illustrated in Figure 6. The sky temperature is based on the 10th percentile of the Galactic foreground brightness distribution and toward the zenith under dry conditions (PWV = 5mm) which should be available about 10% of the time at the SKA1-Mid site. A table providing the values plotted in Figure 6 is included in Appendix A as Table 10.

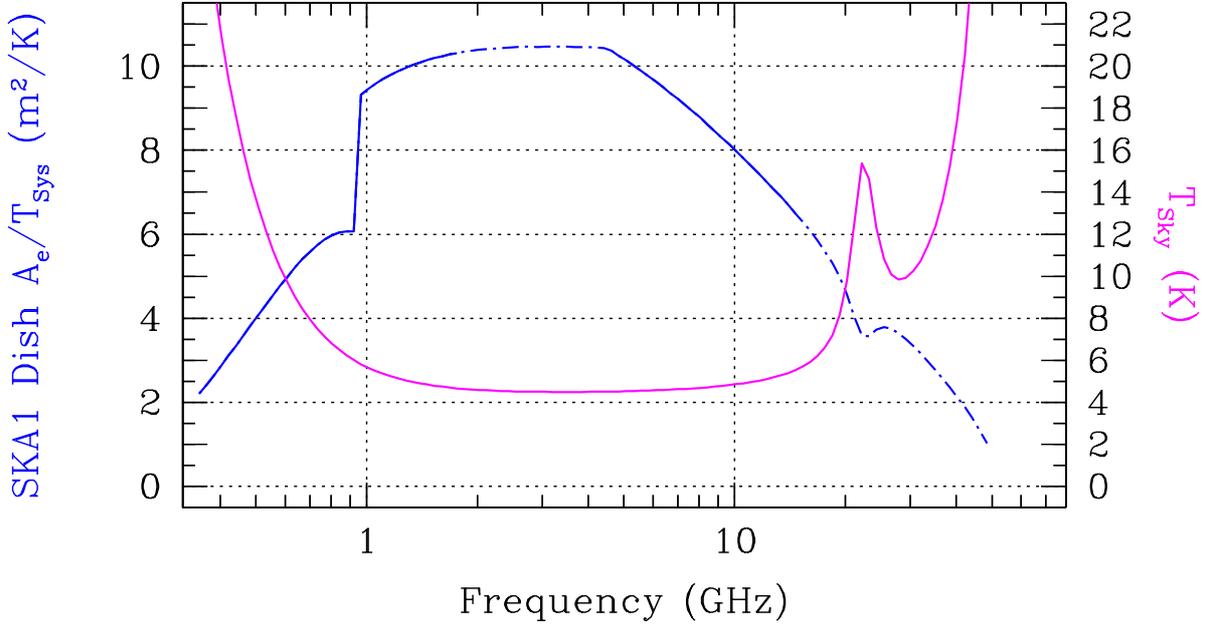

**Figure 6.** Current sensitivity model of a single SKA1 dish at elevations near zenith in a direction well away from the Galactic plane (the 10th percentile) and under dry conditions (PWV = 5 mm). Solid lines are used to indicate the bands that are part of the currently planned deployment.

### 2.1.2 Sensitivity of a MeerKAT Dish

A similar model to the one described above for the SKA dishes can be defined for the MeerKAT dishes. The feed and diffraction efficiencies are modelled as,

$$\eta_F = 0.80 - 0.04 \,|\log_{10}(\nu_{GHz})|$$
and
$$\eta_D = 1 - 20\,(\lambda/D)^{3/2}.$$

The phase efficiency is calculated as before, but the RMS surface errors of the primary and secondary reflector surfaces are assumed to be, $\varepsilon_p = 480$ μm and $\varepsilon_s = 265$ μm, to be consistent with a total surface accuracy of about 600 μm RMS (MeerKAT2016). We assume $T_{spl} = 4$ K and model $T_{rcv}$ as:

Band UHF (0.58 – 1.02 GHz):    $T_{rcv} = 11 - 4.5\,(\nu_{GHz} - 0.58)$ K
Band L (0.9 – 1.67 GHz):    $T_{rcv} = 7.5 + 6.8\,(|\nu_{GHz} - 1.65|)^{3/2}$ K
Band S (1.65 – 3.05 GHz):    $T_{rcv} = 7.5$ K.

This model has been calibrated against the on-sky L-Band measurements reported in MeerKAT2016. The UHF and S-Band models should be regarded as preliminary.

### 2.1.3 Combined SKA1-Mid Sensitivity

The combined sensitivity of SKA1-Mid, including both the 133 SKA1 15 m dishes as well as the 64 MeerKAT 13.5 m dishes is given by the sum of $A_{eff}/T_{sys}$ over all dishes and is listed in Table 10. As noted above, the MeerKAT dishes are only assumed to contribute in the UHF, L and S bands. This combined sensitivity will be relevant for imaging applications as discussed further in Section 8.

## 2.2 SKA1-Low

The antenna that will be used within the SKA1-Low stations is likely to be the SKALA4.1 design, for which prototypes are currently being characterised on site (Waterston et al. 2019). As a placeholder, we will consider the sensitivity parameters provided in de Lera Acedo et al. (2017) for what is termed the SKALA4_BT16 design as deployed within stations having a 40m diameter and each consisting of 256 randomly distributed antennas. The values that will be plotted are for 512 stations and are averaged over all solid angles within 45deg of zenith and are referenced to an assumed Galactic foreground contribution with, $T_{408}$ = 20 K. As noted previously, this corresponds to a value between the $10^{th}$ and $50^{th}$ percentile of the all-sky distribution and would apply to directions well away from the Galactic plane.

## 2.3 SKA2 Sensitivity

While the detailed scope of the full SKA (often call SKA2) is still to be determined, we provide performance projections/aspirations based on an assumed deployment of 2000 SKA1 15-m dishes that are equipped with two Phased Array Feeds to cover 0.35 – 1.67 GHz and octave band feeds between 1.65 and 50 GHz, together with 4880 SKA1 40m diameter low frequency stations that provide performance between 50 and 350 MHz.

# 3 Sensitivity Calculations for other Radio Telescopes

In this section the sensitivities of other telescopes which have been operating long enough to have well characterised on-sky performance or have well determined anticipated performance are presented. In updates to this paper actual performance numbers will be included where possible, as well as measured performance numbers for the other SKA precursor telescopes, MeerKAT and ASKAP.

## 3.1 Jansky Very Large Array (JVLA)

The System Equivalent Flux Density (SEFD) of the recently upgraded Jansky Very Large Array (JVLA or simply VLA) antennas is provided in the Observational Status Summary as function of frequency in VLA2018. The SEFD is related to $A_{eff}/T_{sys}$ by, SEFD = $2 k_B T_{sys}/A_{eff}$, for the Boltzmann constant $k_B$. The high frequency parameters are appropriate for dry conditions and are corrected for atmospheric opacity at the zenith.

## 3.2 ALMA

The lowest frequency bands of the Atacama Large Millimeter/Submillimeter Array (ALMA) overlap with the range accessible to the SKA1. The ALMA interferometric array consists of 50 dishes of 12 m diameter and 12 dishes of 7 m diameter. The sensitivity is calculated following ALMA2017 (ALMA Cycle 5 Technical Handbook), using equations 9.7 and 9.8, together with Table 9.2 from that reference. As previously, we calculate the sky temperature and opacity for the ALMA site, at an assumed elevation of 5100m, ambient temperature of 268 K and atmospheric pressure of 555 mbar. The fiducial water vapour levels for the calculation were taken to be PWV = 0.5, 2 and 5mm, which correspond approximately to the $10^{th}$, $50^{th}$ and $90^{th}$ percentiles of the PWV distribution at the ALMA site. The atmospheric emission and absorption for the ALMA site are shown in Figure 7. The ALMA Band 3 (84-116 GHz) and Band 1 (35-50 GHz) frequencies are indicated with solid lines in the plots, since Band 3 is already available and Band 1 deployment is anticipated in the near future. The ALMA Band 2 frequency range is indicated with a dot-dash line in the plots, since it has not yet been funded for deployment.

## 3.3 upgraded Giant Meterwave Radio Telescope (uGMRT)

The anticipated performance of the upgraded GMRT array, consisting of 30 dishes of 45m diameter has been provided by Lal, Gupta and Chandra (2016). Four feed systems together provide frequency coverage from 120 – 1390 MHz.

## 3.4 LOw Frequency ARray (LOFAR)

The theoretical LOFAR sensitivity has been calculated following the prescription given by Nijboer, Pandey-Pommier and De Bruyn (2009).

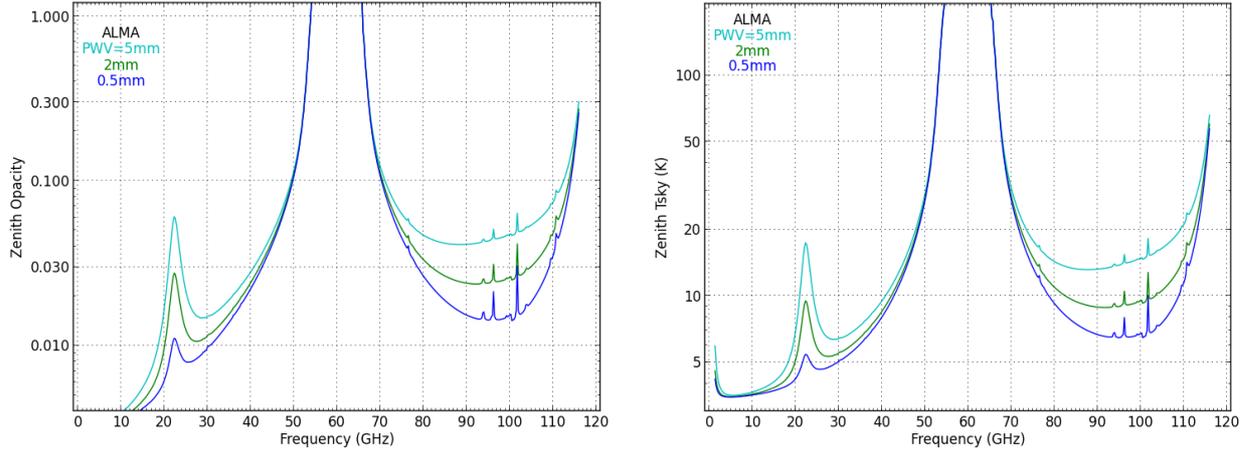

**Figure 7. (Left)** Zenith $T_{sky}$ at the ALMA site for PWV = 0.5, 2 and 5 mm contributions to the atmospheric emission and the $10^{th}$, $50^{th}$ and $90^{th}$ percentiles of the Galactic foreground brightness. Atmospheric emission is based on the ATM code (Pardo et al. 2001) as embedded in CASA. **(Right)** Zenith opacity at the ALMA site for PWV = 0.5, 2 and 5 mm contributions to the atmospheric emission.

# 4 Sensitivity Comparison

The sensitivity of the SKA is compared to various existing and planned facilities in Figure 8 based on the values documented in Sections 2 and 3 above.

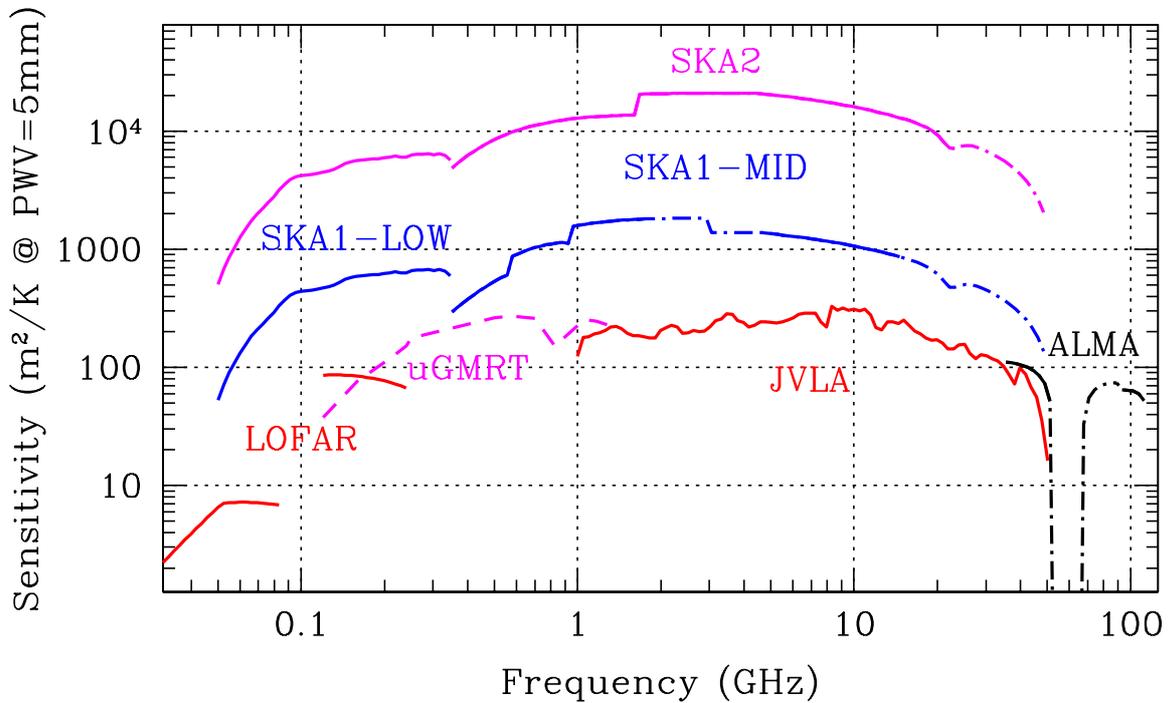

**Figure 8**. Sensitivity comparison of some existing and planned facilities. For SKA, the feed systems that are not yet planned for deployment are indicated by the dot-dashed line. Dry conditions (PWV ≈ 5 mm) are assumed for the SKA and VLA sites while the same PWV = 5 mm corresponds to poor conditions for the ALMA site.

# 5 Survey Speed Comparison

The generic Survey Speed Figure of Merit, SSFOM (aka the survey speed), can be expressed as the product of sensitivity squared with the noise effective Field of View,

$$\mathrm{SSFOM} = S_M^2\, \mathrm{FoV}_{\mathrm{eff}}.$$

The noise effective $\mathrm{FoV}_{\mathrm{eff}}$ is provided by the integral of the square of the normalised primary beam pattern. For the case of a dish with a 10 dB edge taper to the illumination pattern this is given approximately by

$$\mathrm{FoV}_{\mathrm{eff}} = 2340\, (\lambda/D)^2\ \mathrm{deg}^2,$$

in terms of the observing wavelength, $\lambda$, and dish diameter, $D$.

For the case of a uniformly illuminated aperture (such as an aperture array that is beam-formed for maximum sensitivity) one has instead,

$$\mathrm{FoV}_{\mathrm{eff}} = 1920\, (\lambda/D)^2\ \mathrm{deg}^2.$$

The survey speed of the SKA is compared to various existing and planned facilities in Figure 9 based on the sensitivity values documented in Sections 3 and 4 above, together with the dish or station diameters noted there. In those case where multiple dish diameters contribute to the array, only the largest is used in the calculation. The $\mathrm{FoV}_{\mathrm{eff}}$ for the special case of Phased Array Feeds (assumed in the SKA2 deployment) is described in SKAO-ST et al. (2016).

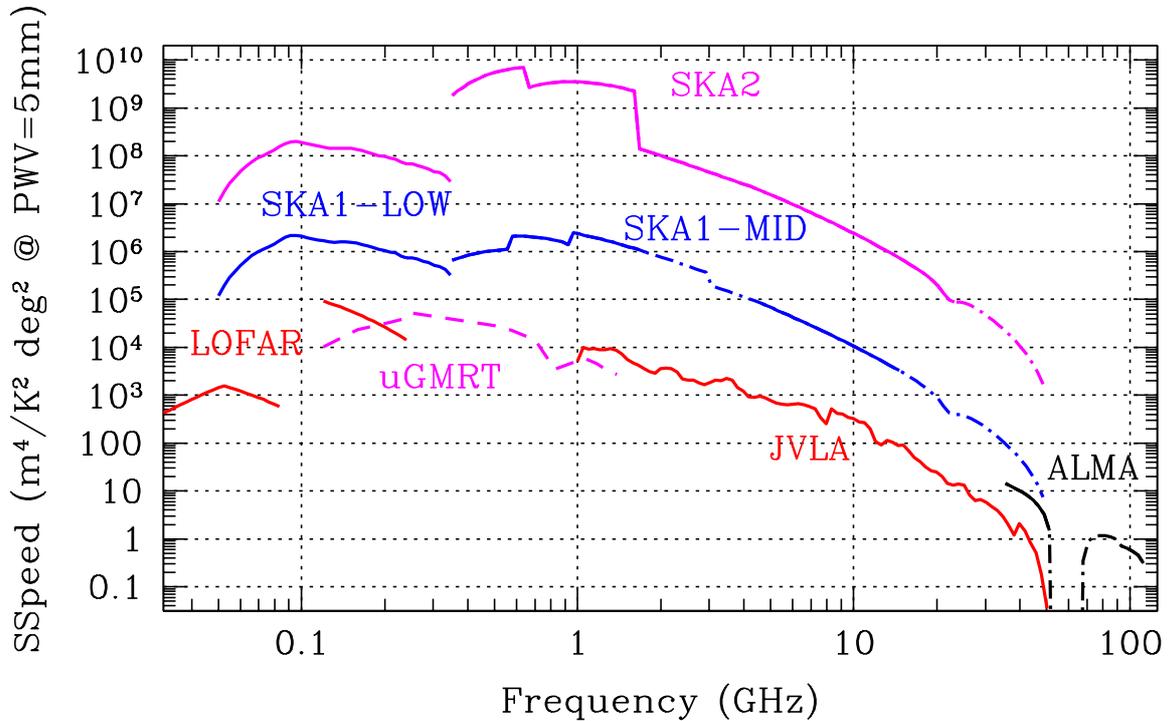

**Figure 9.** Survey speed comparison of some existing and planned facilities. For SKA, the feed systems that are not yet planned for deployment are indicated by the dot-dashed line. Dry conditions (PWV ≈ 5 mm) are assumed for the SKA and VLA sites while the same PWV = 5 mm corresponds to poor conditions for the ALMA site.

# 6   Array Sensitivity

The information presented in the preceding sections, along with the array configurations, are used to calculate both the expected sensitivity and the range of angular scales over which the calculation is valid, for some representative observations. While Braun et al. (2017) considers a number of imaging attributes, here we focus on the relative array sensitivity determined through imaging simulations, compared to the "natural" sensitivity as calculated in Section 4. These results demonstrate how the logarithmic spacing of the array configurations provides a fairly uniform sensitivity to a wide range of angular scales. Values for the imaging sensitivity are presented in Section 8.

The basic observing mode that is employed for an imaging experiment with an array of dishes or aperture array stations varies between one (or more) short snap-shots to a more extensive sky-tracking observation. Since the visibility sampling is improved with long duration tracks, we will consider, as one extreme, the longest practical tracks consistent with good performance for the type of receptor being employed. This corresponds to 8 hour tracks that extend from – 4 h to + 4 h of Hour Angle for the dish arrays and 4 hour tracks, extending from – 2 h to + 2 h for the aperture arrays. While the character and quality of the visibility coverage are also influenced to some degree by the Declination of the field being observed, we will concentrate on the middle of the accessible Dec. range near – 30° as an illustrative example. Generally, the coverage will become even more symmetric and complete for more negative Declinations, while Declinations near 0 are the most limited. At the other extreme from the long tracks, we will consider the case of a single snap-shot observation near transit.

Further, we will consider both "spectral line" and "continuum" observing modes, with "spectral line" providing essentially monochromatic visibilities, and with "continuum" being defined as constituting a 30% fractional bandwidth that is simulated with 30 spectral channels each of 1% width. There are various reasons why mixing larger fractional bandwidths into a single image may not be very meaningful, and we consider this question in more detail in Section 7 below. It should be noted that under many circumstances it will be possible to acquire more than 30% fractional bandwidth simultaneously and the larger the instantaneous bandwidth, the greater the scientific utility in general.

The results are presented in Figure 10 and Figure 11 and Table 5. These results indicate that the image noise relative to the natural array sensitivity is about 2-2.5 over a wide range of angular scales for both continuum images and spectral line cubes.

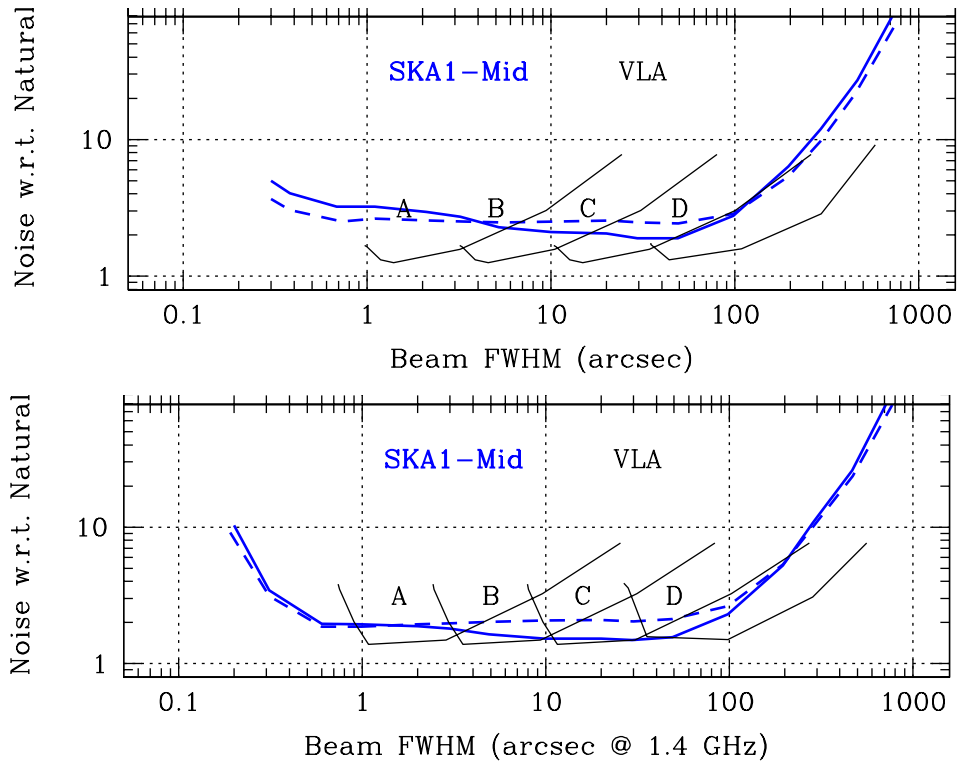

**Figure 10. (Top)** SKA1-Mid monochromatic ("spectral line") image performance as function of angular scale for a nominal frequency of 1.4 GHz. The panel shows the image sensitivity relative to the natural array sensitivity for images with a dirty beam as Gaussian as possible, for an 8h tracking observation. **(Bottom)** SKA1-Mid broadband image performance as function of angular scale for a nominal frequency of 1.4 GHz. The lower panel provides the image sensitivity relative to the natural array sensitivity for images with a dirty beam as Gaussian as possible, for an 8h tracking observation. In both panels the solid blue curves include both the SKA and MeerKAT dishes, while the dashed curves include only the SKA dishes. The performance of the VLA is shown in black lines for each array configuration.

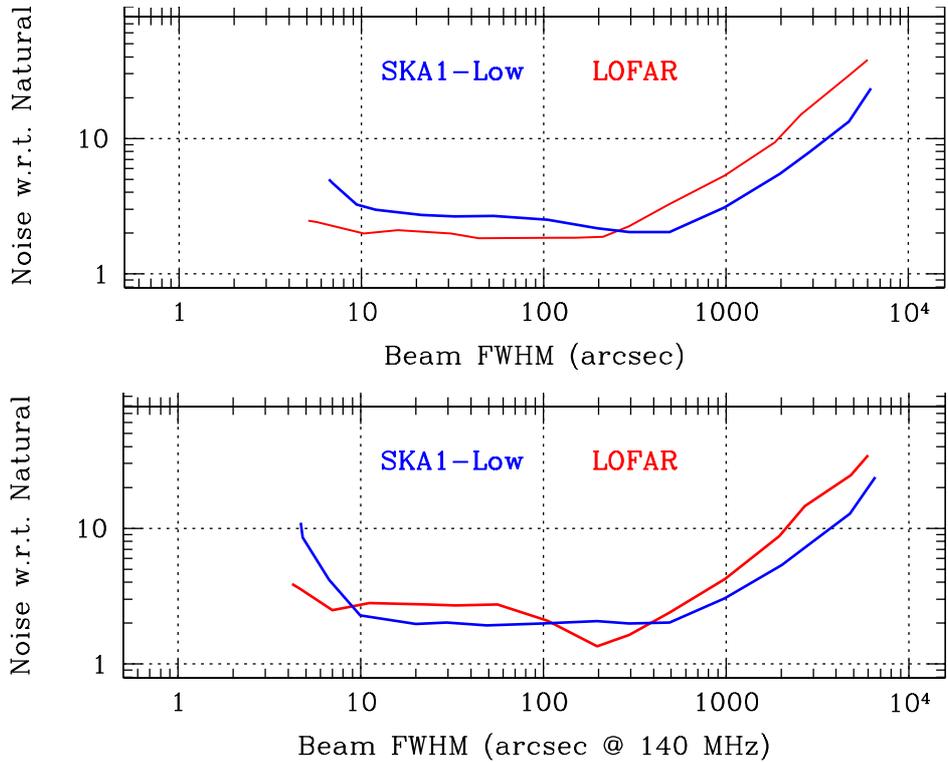

**Figure 11. (Top)** SKA1-Low monochromatic ("spectral line") image performance as function of angular scale for a nominal frequency of 140 MHz. The panel shows the image sensitivity relative to the natural array sensitivity for images with a dirty beam as Gaussian as possible, for a 4h tracking observation. **(Bottom)** SKA1-Low broadband image performance as function of angular scale for a nominal frequency of 140 MHz. The panel shows the image sensitivity relative to the natural array sensitivity for images with a dirty beam as Gaussian as possible, for a 4h tracking observation. The performance of LOFAR is shown in red lines.

**Table 5.** The imaging sensitivity degradation relative to the natural array sensitivity provided by the SKA1-Low and SKA1-Mid array configurations as plotted in Figure 10 and Figure 11 is listed as a function of the FWHM of the most Gaussian possible dirty beam provided by a (4h for Low and 8h for Mid) tracking observation at a reference frequency (140 MHz for Low and 1.4 GHz for Mid). Image attributes at any other frequency can be inferred by a linear scaling to higher or lower angular scales for higher or lower frequencies. The Sens*/Nat columns for SKA1-Mid are relevant for frequencies at which the MeerKAT dishes do not contribute to the array.

| Low Line | | Low Cont | | Mid Line | | | Mid Cont | | |
|---|---|---|---|---|---|---|---|---|---|
| **FWHM (arcsec)** | Sens /Nat | **FWHM (arcsec)** | Sens /Nat | **FWHM (arcsec)** | Sens /Nat | Sens* /Nat | **FWHM (arcsec)** | Sens /Nat | Sens* /Nat |
| **6.6** | 5.0 | **4.7** | 11.0 | **0.3** | 5.0 | 3.7 | **0.2** | 10.3 | 9.2 |
| **7** | 4.7 | **4.8** | 8.6 | **0.4** | 4.1 | 3.1 | **0.3** | 3.4 | 3.1 |
| **9.4** | 3.2 | **6.7** | 4.2 | **0.7** | 3.2 | 2.5 | **0.6** | 2.0 | 1.9 |
| **12** | 3.0 | **10** | 2.3 | **1** | 3.2 | 2.6 | **1** | 1.9 | 1.9 |
| **21** | 2.7 | **20** | 2.0 | **2** | 2.9 | 2.6 | **2** | 1.9 | 1.9 |
| **32** | 2.7 | **30** | 2.0 | **3** | 2.7 | 2.5 | **3** | 1.8 | 2.0 |
| **53** | 2.7 | **49** | 1.9 | **5** | 2.3 | 2.5 | **5** | 1.6 | 2.0 |
| **105** | 2.5 | **98** | 2.0 | **10** | 2.1 | 2.5 | **10** | 1.5 | 2.1 |
| **197** | 2.2 | **196** | 2.1 | **20** | 2.1 | 2.6 | **20** | 1.5 | 2.1 |
| **295** | 2.0 | **295** | 2.0 | **30** | 1.9 | 2.5 | **30** | 1.5 | 2.0 |
| **491** | 2.0 | **491** | 2.0 | **49** | 1.9 | 2.4 | **49** | 1.6 | 2.1 |
| **983** | 3.1 | **983** | 3.0 | **98** | 2.8 | 2.9 | **98** | 2.3 | 2.6 |
| **1980** | 5.5 | **2020** | 5.4 | **196** | 6.3 | 5.4 | **196** | 5.2 | 5.3 |
| **2875** | 8.0 | **2874** | 7.7 | **293** | 11.9 | 9.7 | **284** | 10.7 | 10.0 |
| **4720** | 13.4 | **4750** | 12.8 | **465** | 27.1 | 21.5 | **468** | 26.0 | 23.6 |
| **6260** | 23.4 | **6550** | 23.8 | **720** | 80.8 | 66.0 | **768** | 97.2 | 78.6 |

# 7 Bandwidth Considerations for Continuum Observing

When quantifying the scientific performance of a radio telescope for the study of broad-band continuum phenomena, there are a number of issues that need to be considered. First is the intrinsic SED (spectral energy distribution) of the target source population. Only a small fraction of the faint extragalactic source population can be described by a relatively flat spectral index, $\alpha \approx 0$, where $S_\nu = S_0 (\nu/\nu_0)^\alpha$ [cf. Figure 7 of Mancuso et al. (2017)]. And even when this is the case, it is unlikely to be a good representation of the spectrum over a large fractional bandwidth. Other source populations vary between the ultra-steep negative spectra ($\alpha = -1$ to $-2$) of diffuse galaxy cluster sources as well as much of the pulsar population, to the ultra-steep positive spectra ($\alpha = +3$ to $+4$) for the modified black-body of dust emission sources. To document the scientific utility of broad-band observations we plot the signal-to-noise ratio for sources of varying spectral index as a function of the fractional bandwidth employed for the observation in Figure 12.

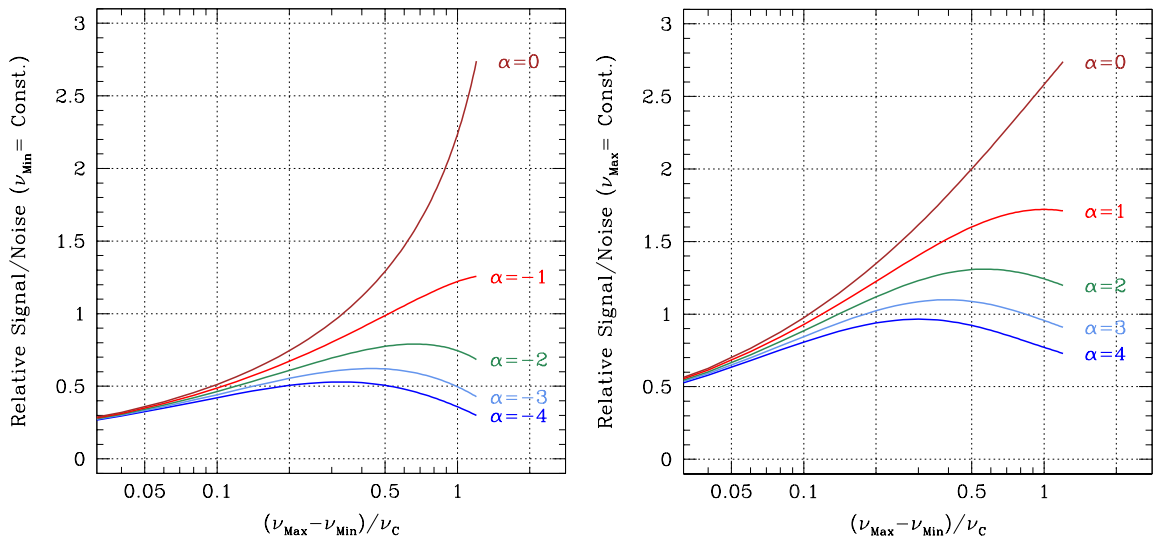

**Figure 12.** The relative signal-to-noise ratio as function of fractional bandwidth for varying spectral index of the source population. For negative spectral indices (left) the minimum observing frequency is kept fixed, while for positive spectral indices (right) the maximum observing frequency is kept fixed.

Since the largest signal contribution comes from the minimum observing frequency (for $\alpha<0$) or the maximum observing frequency (for $\alpha>0$) we keep this quantity fixed while varying the fractional bandwidth, $(\nu_{Max} - \nu_{Min})/\nu_C$, in the plots. The curve for $\alpha=0$ describes the typical assumption of an improvement in signal-to-noise that is proportional to the square root of bandwidth. As is apparent from the plots, there are significant departures from the square root bandwidth approximation for even modest values of the fractional bandwidth of 10 – 20%. For the more extreme spectral indices, the signal-to-noise ratio even declines for fractional bandwidths exceeding about 30%, since more noise than signal is then being averaged in the outcome. It is apparent from this demonstration that the practice of quoting continuum sensitivities based on a large fractional bandwidth is misleading.

A second factor to consider when documenting the broad-band performance is whether a single telescope pointing is being utilised for the observation of interest, or whether a large area of the sky is being studied that requires multiple telescope pointings (as determined by the highest observing frequency under consideration). In the event that multiple telescope pointings are required, one must take account of the different noise effective field-of-view that is available as function of frequency. Since the FoV scales as $\nu^{-2}$ (as shown in Section 5), the density of pointings that is required to achieve a uniform net sensitivity on the sky is determined by the highest observing frequency. Consequently, the degree of pointing over-sampling increases linearly in each direction toward lower frequencies, with the net effect that the survey sensitivity increases as $\nu^{-1}$, for the case of an intrinsic sensitivity that is constant with frequency.

Finally, one must consider the intrinsic telescope variation of sensitivity with frequency itself, such as shown in Figure 8 and documented in Table 3 and Table 5. For large fractional bandwidth, this variation can also become significant.

In view of all these considerations, we prefer to limit performance projections of broad-band continuum observations to no more than about 30% fractional bandwidth. However, as shown in Figure 12, even this limit will be overly optimistic in predicting the actual signal-to-noise by up to a factor of two for some plausible source populations.

# 8   Imaging Sensitivity

Given the array sensitivity documented in Section 4 together with the sensitivity as function of angular scale in Section 6 it is straightforward to provide estimates of the image sensitivity for specific observations. The SEFD (System Equivalent Flux Density) is given by,

$$\text{SEFD} = 2\, k_B\, T_{sys}/A_{eff},$$

where $k_B$ is the Boltzmann constant.

The image noise is given by,

$$\sigma = S_D\, \text{SEFD} / [\eta_S\, (n_{pol}\, \Delta\nu\, \Delta\tau)^{1/2}],$$

where $S_D$ is a degradation factor relative to the natural array sensitivity for the specific target Gaussian FWHM resolution of the image. As discussed in Section 6 (Table 5), this degradation factor is approximately $S_D \approx 2$ for a wide range of intermediate angular scales and climbs to larger values for extreme angular scales. The factor $\eta_S$ is a system efficiency that takes account of the finite correlator efficiency and other forms of incoherence and is assumed to be $\eta_S = 0.9$ (e.g., Dewdney et al. 2013). The factor $n_{pol}$ is the number of contributing polarisations and for SKA1 with orthogonal linear polarisations is $n_{pol} = 2$. The contributing bandwidth is $\Delta\nu$ and the integration time is $\Delta\tau$. For illustration, we will consider the cases of a spectral line observation with fractional bandwidth of $\Delta\nu/\nu_c = 10^{-4}$ and a continuum observation with fractional bandwidth of $\Delta\nu/\nu_c \approx 0.3$, together with an integration time $\Delta\tau = 1$ hour.

The imaging sensitivity for SKA1-Low and SKA1-Mid is listed in Table 6 and Table 7, based on an assumed degradation factor (Table 5) of $S_D = 2.5$ for the noise in a line image slice, $\sigma_L$, and $S_D = 2$ for the noise in a continuum image, $\sigma_C$. The range of target Gaussian FWHM beam sizes over which this is assumed to apply is given by $\theta_{min}$ to $\theta_{max}$. The beam sizes outside of this range which lead to a doubling of the RMS noise (to $S_D = 4$) are indicated by $\theta'_{min}$ and $\theta'_{max}$. The spectral line sensitivity is the average over the frequency range indicated. The continuum sensitivity is estimated from the square root of the sum of the squared sensitivities over the indicated range. For SKA1-Low, we include an additional column labelled $\sigma_{Conf}(\theta_{min})$, the confusion noise following Condon et al. (2012) that applies at the band centre frequency for the beam size, $\theta_{min}$. As noted in Section 2.1.3, we assume that the MeerKAT dishes will only contribute to the image sensitivity of SKA1-Mid within the frequency ranges of the UHF, L and S bands. Outside of these ranges the array configuration is assumed to include only the SKA dishes.

**Table 6.** Image sensitivity of SKA1-Low within the indicated frequency bands for spectral line observations ($\sigma_L$ for $\Delta\nu/\nu_c = 10^{-4}$) and continuum observations ($\sigma_C$ for $\Delta\nu/\nu_c \approx 0.3$) for an observation of $\Delta\tau = 1$ hour. The range of Gaussian FWHM beam sizes for which the approximate sensitivity value applies is given by $\theta_{min}$ to $\theta_{max}$. The Gaussian FWHM beam sizes at which a doubling of the image noise from this base level is realised are given by $\theta'_{min}$ and $\theta'_{max}$. The anticipated confusion noise at a resolution $\theta_{min}$ and $\nu_c$ is also listed (following Condon et al. 2012).

| $\nu_{min}$ (MHz) | $\nu_c$ (MHz) | $\nu_{max}$ (MHz) | $\sigma_L$ (μJy/Bm) | $\sigma_C$ (μJy/Bm) | $\sigma_{Conf}(\theta_{min})$ (μJy/Bm) | $\theta'_{min}$ (") | $\theta_{min}$ (") | $\theta_{max}$ (") | $\theta'_{max}$ (") |
|---|---|---|---|---|---|---|---|---|---|
| 50 | 60 | 69 | 11050 | 163 | 677 | 16.4 | 23.5 | 1175 | 3290 |
| 69 | 82 | 96 | 3261 | 47 | 183 | 11.9 | 17.0 | 850 | 2379 |
| 96 | 114 | 132 | 1841 | 26 | 50 | 8.6 | 12.3 | 614 | 1719 |
| 132 | 158 | 183 | 1258 | 18 | 13 | 6.2 | 8.9 | 444 | 1244 |
| 183 | 218 | 253 | 973 | 14 | 4 | 4.5 | 6.4 | 321 | 899 |
| 253 | 302 | 350 | 794 | 11 | 1 | 3.3 | 4.6 | 232 | 650 |

**Table 7.** Image sensitivity of SKA1-Mid within the indicated frequency bands for spectral line observations ($\sigma_L$ for $\Delta\nu/\nu_c = 10^{-4}$) and continuum observations ($\sigma_C$ for $\Delta\nu/\nu_c \approx 0.3$) for an observation of $\Delta\tau = 1$ hour. The range of Gaussian FWHM beam sizes for which the approximate sensitivity value applies is given by $\theta_{min}$ to $\theta_{max}$. The Gaussian FWHM beam sizes at which a doubling of the image noise from this base level is realised are given by $\theta'_{min}$ and $\theta'_{max}$. The grey shading is used to approximately indicate those frequencies that are not yet part of the current deployment plan (ie. between 1.76 and 4.6 GHz as well as above 15.3 GHz).

| $\nu_{min}$ (GHz) | $\nu_c$ (GHz) | $\nu_{max}$ (GHz) | $\sigma_L$ (μJy/Bm) | $\sigma_C$ (μJy/Bm) | $\theta'_{min}$ (") | $\theta_{min}$ (") | $\theta_{max}$ (") | $\theta'_{max}$ (") |
|---|---|---|---|---|---|---|---|---|
| 0.35 | 0.41 | 0.48 | 1176 | 16.8 | 1.015 | 2.031 | 270.8 | 541.6 |
| 0.48 | 0.56 | 0.65 | 560 | 8.1 | 0.745 | 1.489 | 198.6 | 397.2 |
| 0.65 | 0.77 | 0.89 | 303 | 4.4 | 0.546 | 1.092 | 145.6 | 291.2 |
| 0.89 | 1.05 | 1.21 | 186 | 2.7 | 0.400 | 0.801 | 106.8 | 213.5 |
| 1.21 | 1.43 | 1.65 | 137 | 2.0 | 0.294 | 0.587 | 78.3 | 156.6 |
| 1.65 | 1.95 | 2.25 | 113 | 1.6 | 0.215 | 0.431 | 57.4 | 114.9 |
| 2.25 | 2.66 | 3.07 | 99 | 1.4 | 0.158 | 0.316 | 42.1 | 84.2 |
| 3.07 | 3.63 | 4.18 | 109 | 1.6 | 0.116 | 0.232 | 30.9 | 61.8 |
| 4.18 | 4.94 | 5.70 | 95 | 1.4 | 0.085 | 0.170 | 22.7 | 45.3 |
| 5.70 | 6.74 | 7.78 | 89 | 1.3 | 0.062 | 0.125 | 16.6 | 33.2 |
| 7.78 | 9.19 | 10.61 | 85 | 1.2 | 0.046 | 0.091 | 12.2 | 24.4 |
| 10.61 | 12.53 | 14.46 | 85 | 1.2 | 0.034 | 0.067 | 8.9 | 17.9 |
| 14.46 | 17.09 | 19.72 | 91 | 1.3 | 0.025 | 0.049 | 6.6 | 13.1 |
| 19.72 | 23.31 | 26.89 | 116 | 1.7 | 0.018 | 0.036 | 4.8 | 9.6 |
| 26.89 | 31.78 | 36.67 | 121 | 1.8 | 0.013 | 0.026 | 3.5 | 7.0 |
| 36.67 | 43.33 | 50.00 | 209 | 3.2 | 0.010 | 0.019 | 2.6 | 5.2 |

# 9 Non-Imaging Applications

Many applications (e.g., time domain astronomy, VLBI) do not make use of the correlated visibilities, but instead rely on a tied array beam made possible through beamforming. These non-imaging applications make use of the SKA's beamformers, by coherently adding the signals from sub-arrays of many receptors. Both SKA1-Low and -Mid have multiple capabilities in this regard.

The sensitivity and effective survey speed for non-imaging applications differ to the generic array figures of merit for two reasons: (i) non-imaging applications do not use the entire array; (ii) there is a finite number of tied-array beams that can be formed, i.e. only a fraction of the FoV is used. The particular constraints for SKA1-Mid and -Low are outlined below but the procedure is essentially a trade-off between sensitivity (sub-array size, increasing with radius from core), tied-array beam size (decreased with radius from core), number of tied-array beams that the beam-former can produce, and compute processing capabilities (e.g. ability to compensate for less sensitivity with more observing time, $T_{obs}$, and compute processing that scales with $\sim T_{obs}^3$ in the case of a Binary Pulsar Search). The overall effect is that, e.g., the survey speed for non-imaging applications can easily be an order of magnitude less than shown in Figure 8.

For pulsar survey/search purposes there is a beamformer that produces Stokes data, reduced in time and frequency resolution in a configurable manner. SKA1-Mid (SKA1-Low) can produce 1500 x 300-MHz (500 x 100-MHz) tied-array beams of this kind and there is the normal ability to swap number of beams for the bandwidth of each beam, modulo some hardware restrictions.

In addition, both telescopes have a pulsar timing beamformer that produces up to 16 voltage beams (8 on SKA1-Mid Band 5) covering the entire observing band in question. Voltage beams are needed so as to perform a lossless coherent de-dispersion of data from pointed observations of known pulsars (with known dispersion measures). A modified observing mode known as "Dynamic Spectrum Mode" is essentially a use of the pulsar timing infrastructure where no de-dispersion or folding (or optionally any RFI flagging/cleaning) is performed, with the data typically reduced in time and frequency (and bit-depth) resolution to a manageable size for offline searches. This latter mode would be the method for performing very deep targeted searches, rather than all-sky searches. A complete "flow through" approach where no operations are performed on the data permits raw voltages to be recorded, which might be used in special case observations and system maintenance.

Both telescopes also support VLBI observing, which itself also requires voltage beams albeit in a different format. VLBI beams on SKA1-Mid are limited to 4 x 2.5-GHz beams (or some trade-off of this in terms of number of beams and bandwidth per beam); on SKA1-Low the voltage beams come as a trade-off for pulsar timing beams and are limited to 4 x 300-MHz beams.

Key non-imaging capabilities and sensitivities are summarised in Table 8 and Figure 13 and Figure 14.

**Table 8** Pulsar and Fast Radio Burst (FRB) capabilities.

|  | **SKA1-Mid** | **SKA1-Low** |
|---|---|---|
| Pulsar & FRB Search beams | 1500 x 300 MHz | 500 x 100 MHz |
| Pulsar Acceleration Search | +/– 350 m/s/s | +/– 350 m/s/s |
| FRB real-time search | 0-3000 pc/cc | 0-3000 pc/cc |
| Transient voltage buffer size | 9 minutes | 30 seconds |
| Timing precision | < 5ns | < 10 ns |

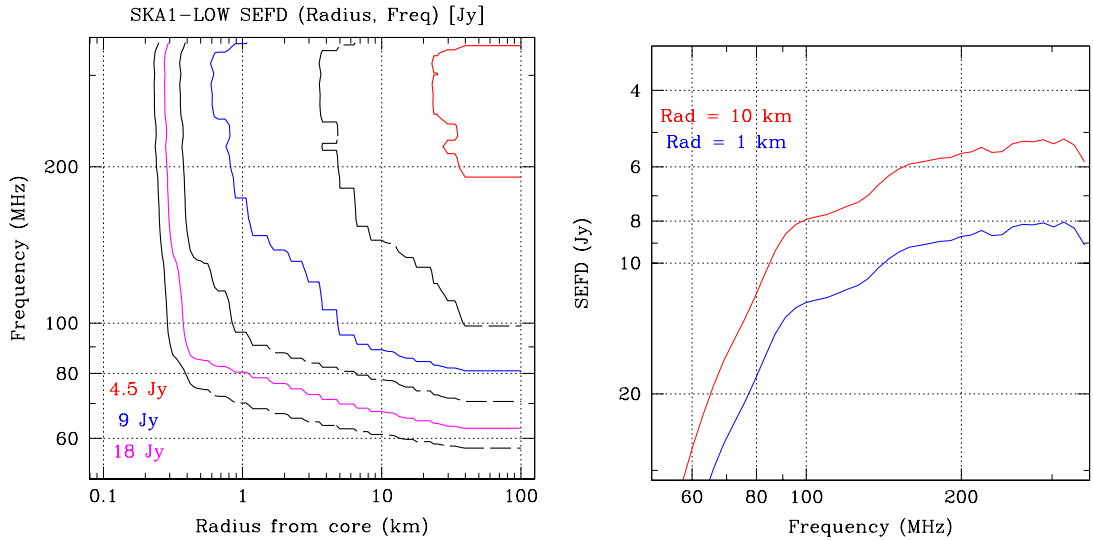

**Figure 13.** The tied array sensitivity of SKA1-Low as function of the maximum radius of included stations (left) as well as the sensitivity as function of frequency for two reference radii (right). The model is appropriate for an average elevation within 45 degrees of zenith in a direction well away from the Galactic plane.

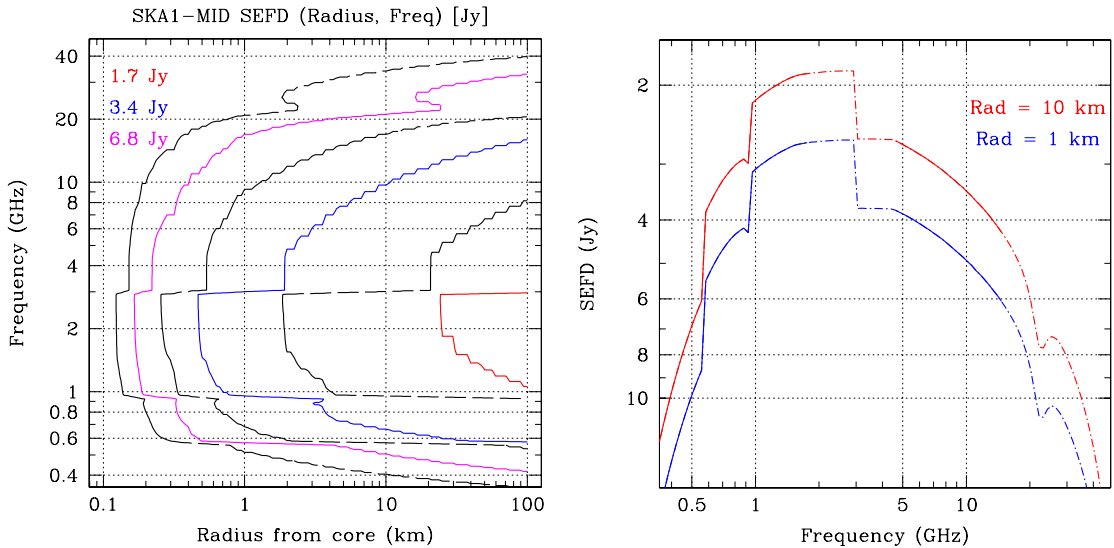

**Figure 14.** The tied array sensitivity of SKA1-Mid as function of the maximum radius of included dishes (left) as well as the sensitivity as function of frequency for two reference radii (right). The model is appropriate for elevations near zenith in a direction well away from the Galactic plane (the 10[th] percentile) and under dry conditions (PWV = 5mm). Solid lines (right) are used to indicate the bands that are part of the currently planned deployment.

**Version History**
1.0 26 December 2019

# Appendix A: Natural Sensitivities

The numerical values shown in several key plots are listed here in Table form for convenience. The individual SKA1-Low station and the combined SKA1-Low array natural sensitivities from Figure 8 are shown in Table 9. Recall, that these are averaged over all solid angles within 45 degrees of zenith. The individual SKA1-Mid dish and the combined SKA1-Mid array natural sensitivities from Figure 6 and Figure 8 are shown in Table 10. The imaging sensitivity provided by the SKA1-Low and SKA1-Mid array configurations as function of scale, as plotted in Figure 10 and Figure 11 have already been listed in Table 5.

**Table 9.** The individual SKA1-Low station and the combined SKA1-Low array natural sensitivities from Figure 8 are listed here as function of frequency. The sensitivity is averaged over solid angles within 45 degrees of zenith.

| Frequency (MHz) | SKA1-Low Station $A_{eff}/T_{sys}$ (m$^2$/K) | SKA1-Low Array $A_{eff}/T_{sys}$ (m$^2$/K) |
|---|---|---|
| 50 | 0.102 | 52.2 |
| 60 | 0.255 | 130.3 |
| 70 | 0.420 | 214.8 |
| 80 | 0.580 | 296.9 |
| 90 | 0.781 | 399.7 |
| 100 | 0.862 | 441.3 |
| 110 | 0.885 | 453.1 |
| 120 | 0.924 | 472.9 |
| 130 | 0.965 | 493.9 |
| 140 | 1.052 | 538.4 |
| 150 | 1.119 | 572.8 |
| 160 | 1.159 | 593.3 |
| 170 | 1.171 | 599.7 |
| 180 | 1.190 | 609.1 |
| 190 | 1.195 | 612.0 |
| 200 | 1.223 | 626.4 |
| 210 | 1.235 | 632.4 |
| 220 | 1.263 | 646.5 |
| 230 | 1.226 | 627.5 |
| 240 | 1.234 | 631.7 |
| 250 | 1.282 | 656.4 |
| 260 | 1.304 | 667.9 |
| 270 | 1.297 | 664.2 |
| 280 | 1.306 | 668.8 |
| 290 | 1.314 | 672.5 |
| 300 | 1.289 | 660.0 |
| 310 | 1.294 | 662.4 |
| 320 | 1.327 | 679.4 |
| 330 | 1.286 | 658.6 |
| 340 | 1.204 | 616.7 |
| 350 | 1.156 | 591.8 |

**Table 10.** The individual SKA1-Mid dish and the combined SKA1-Mid array natural sensitivities from Figure 6 and Figure **8** are listed here as function of frequency. The grey shading is used to indicate frequencies that are not yet part of the current deployment plan. Note that the MeerKAT dishes are only assumed to contribute to the array sensitivity within the UHF, L and S Bands as described in Section 2.1.3.

| Frequency (GHz) | SKA1-Mid Dish $A_{eff}/T_{sys}$ (m²/K) | SKA1-Mid Array $A_{eff}/T_{sys}$ (m²/K) |
|---|---|---|
| **0.3508** | 2.208 | 293.7 |
| **0.3673** | 2.424 | 322.3 |
| **0.3846** | 2.648 | 352.2 |
| **0.4027** | 2.879 | 382.9 |
| **0.4217** | 3.114 | 414.2 |
| **0.4416** | 3.353 | 445.9 |
| **0.4624** | 3.594 | 478.1 |
| **0.4842** | 3.837 | 510.4 |
| **0.507** | 4.078 | 542.4 |
| **0.5309** | 4.316 | 574 |
| **0.5559** | 4.55 | 605.1 |
| **0.5821** | 4.779 | 877.1 |
| **0.6095** | 5 | 918.2 |
| **0.6383** | 5.209 | 957.4 |
| **0.6683** | 5.404 | 994.4 |
| **0.6998** | 5.581 | 1029 |
| **0.7328** | 5.737 | 1060 |
| **0.7674** | 5.869 | 1088 |
| **0.8035** | 5.971 | 1112 |
| **0.8414** | 6.037 | 1130 |
| **0.881** | 6.072 | 1144 |
| **0.9226** | 6.071 | 1124 |
| **0.9661** | 9.324 | 1567 |
| **1.012** | 9.47 | 1597 |
| **1.059** | 9.59 | 1623 |
| **1.109** | 9.698 | 1647 |
| **1.161** | 9.796 | 1671 |
| **1.216** | 9.883 | 1693 |
| **1.274** | 9.961 | 1713 |
| **1.334** | 10.03 | 1733 |
| **1.396** | 10.09 | 1751 |
| **1.462** | 10.15 | 1767 |
| **1.531** | 10.2 | 1781 |
| **1.603** | 10.24 | 1792 |
| **1.679** | 10.28 | 1799 |
| **1.758** | 10.31 | 1804 |
| **1.841** | 10.34 | 1810 |
| **1.928** | 10.37 | 1814 |
| **2.018** | 10.39 | 1818 |
| **2.113** | 10.41 | 1821 |
| **2.213** | 10.42 | 1824 |
| **2.317** | 10.44 | 1826 |
| **2.427** | 10.45 | 1828 |
| **2.541** | 10.46 | 1829 |
| **2.661** | 10.46 | 1830 |
| **2.786** | 10.47 | 1831 |
| **2.917** | 10.46 | 1830 |
| **3.055** | 10.46 | 1392 |
| **3.199** | 10.46 | 1392 |
| **3.35** | 10.46 | 1392 |
| **3.508** | 10.46 | 1391 |
| **3.673** | 10.46 | 1391 |
| **3.846** | 10.45 | 1390 |
| **4.027** | 10.45 | 1390 |
| **4.217** | 10.44 | 1389 |
| **4.416** | 10.43 | 1388 |
| **4.624** | 10.36 | 1378 |
| **4.842** | 10.25 | 1363 |
| **5.07** | 10.13 | 1348 |
| **5.309** | 10.01 | 1332 |
| **5.559** | 9.891 | 1315 |
| **5.821** | 9.764 | 1299 |
| **6.095** | 9.633 | 1281 |
| **6.383** | 9.5 | 1263 |
| **6.683** | 9.362 | 1245 |
| **6.998** | 9.222 | 1226 |
| **7.328** | 9.078 | 1207 |
| **7.674** | 8.931 | 1188 |
| **8.035** | 8.78 | 1168 |
| **8.414** | 8.617 | 1146 |
| **8.81** | 8.461 | 1125 |
| **9.226** | 8.299 | 1104 |
| **9.661** | 8.138 | 1082 |
| **10.12** | 7.97 | 1060 |
| **10.59** | 7.801 | 1038 |
| **11.09** | 7.617 | 1013 |
| **11.61** | 7.446 | 990.3 |
| **12.16** | 7.267 | 966.5 |
| **12.74** | 7.084 | 942.2 |
| **13.34** | 6.89 | 916.4 |
| **13.96** | 6.7 | 891.1 |
| **14.62** | 6.497 | 864.1 |
| **15.31** | 6.295 | 837.3 |
| **16.03** | 6.078 | 808.4 |
| **16.79** | 5.85 | 778 |
| **17.58** | 5.597 | 744.4 |
| **18.41** | 5.317 | 707.2 |
| **19.28** | 4.985 | 663 |
| **20.18** | 4.564 | 607 |
| **21.13** | 4.043 | 537.7 |
| **22.13** | 3.587 | 477.1 |
| **23.17** | 3.574 | 475.3 |
| **24.27** | 3.726 | 495.6 |
| **25.41** | 3.791 | 504.3 |
| **26.61** | 3.751 | 498.9 |
| **27.86** | 3.648 | 485.2 |
| **29.17** | 3.507 | 466.4 |
| **30.55** | 3.337 | 443.9 |
| **31.99** | 3.163 | 420.6 |
| **33.5**  | 2.971 | 395.1 |
| **35.08** | 2.77  | 368.4 |
| **36.73** | 2.562 | 340.8 |
| **38.46** | 2.348 | 312.2 |
| **40.27** | 2.122 | 282.2 |
| **42.17** | 1.883 | 250.4 |
| **44.16** | 1.626 | 216.2 |
| **46.24** | 1.345 | 178.9 |
| **48.42** | 1.027 | 136.6 |